# Reconstructing Speech Stimuli From Human Auditory Cortex Activity Using a WaveNet-like Network

*Ran Wang[1], Yao Wang[1], Adeen Flinker[2]*

[1] Dept. Electrical & Computer Engineering, Tandon School of Engineering, New York University, Brooklyn, New York, USA
[2] Langone School of Medicine, New York University, New York, New York, USA
{rw1691, yw523}@nyu.edu, Adeen.Flinker@nyumc.org

**Abstract**—The superior temporal gyrus (STG) region of cortex critically contributes to speech recognition. In this work, we show that a proposed deep network inspired by WaveNet, trained with limited available data, is able to reconstruct speech stimuli from STG intracranial recordings. We further investigate the impulse response of the fitted model for each recording electrode and observe phoneme level temporospectral tuning properties in some recorded area. This discovery is consistent with previous studies implicating the posterior STG (pSTG) in a phonetic representation of speech and provides detailed acoustic features that certain electrode sites possibly extract during speech recognition.

## I. INTRODUCTION

Research studies on the superior temporal gyrus (STG) cortex area have shown that this area plays an important role in words and sentence recognition on a phonetic and prelexical stage [1]–[9]. It exhibits selectivity to distinct clusters of speech phonetic spectra-temporal features [6], [10]. Linear regression models have been applied to reconstruct speech stimuli from intracranial Electrocorticographic (ECoG) recordings of STG to quantitatively demonstrate STG cortical representations [11]. However, the recovered speech stimuli from STG brain signals has shown limited intelligibility. Such attempts can provide research tools of finer scope to study STG and other cortices.

Many BCI systems have been built based on ECoG recordings. A multi-scale convolutional network approach has been applied to predict upcoming high-density ECoG signals in an animal model of epilepsy in order to detect seizure onset [12]. Wang el al. applied LSTM to predict human arm movement from brain activity [13]. A hierarchical CNN-RNN model was introduced in [14] to decode finger trajectory. To decode recognizable speech requires not only recovering complex acoustic features for the speech waveform, but also uncovering the non-linear relationship between the ECoG signal and the acoustic features. The success of deep neural network in other neural signal processing problems motivated us to investigate deep models decoding speech stimuli from the recorded STG signals.

The success of deep learning essentially relies on a dataset of sufficient training samples for the target problem. Designing a deep network with limited training data is challenging. However, obtaining a big dataset is not feasible with ECoG recordings. In this study, we collected ECoG data from two subjects listening to five-minutes of English words presented auditorily. Since orientation and offset of the implemented ECoG grid varied across patients, obtained signals were different for the two subjects, making combining data from both subjects very challenging. Therefore, we evaluated two datasets separately. These datasets are much smaller than typical deep learning applications. To prevent the model from overfitting while keeping as many fitting parameters as possible to enhance the model's expressive power, we adapted a WaveNet model [15] with high parameter efficiency for the speech decoding task.

We compared decoding results of the adapted WaveNet model with linear regression and ResNet model. Both quantitative and qualitative comparisons show that WaveNet generates reconstructions closer to the original stimuli and is a novel promising approach to generate intelligible speech from brain signals. To further validate the effectiveness of the model, we investigated the response of the fitted network to the impulse signal from each electrode input. Some of the discovered impulse responses resemble recognizable phonemes and provide an estimate of an encoding model [16]. This finding is consistent with the phonetic selectivity of STG reported in [6] and our results suggest specific tuning for individual phonemes in certain electrode sites.

## II. METHOD

With the constraint of a small dataset on which trained models are easily overfitted, a trade-off between expressive power and generalization performance must be considered. To achieve the same representation power, the model with higher parameter efficiency should have fewer trainable parameters and thus is less likely to overfit.

Based on the above consideration, we modified the WaveNet model, the state-of-art model for waveform generation, to recover speech spectrograms from recorded ECoG responses. The original WaveNet takes noise samples as input and sequentially generates random-meaning audio. If trained on natural speech dataset, the model is able to produce real sounding speech.

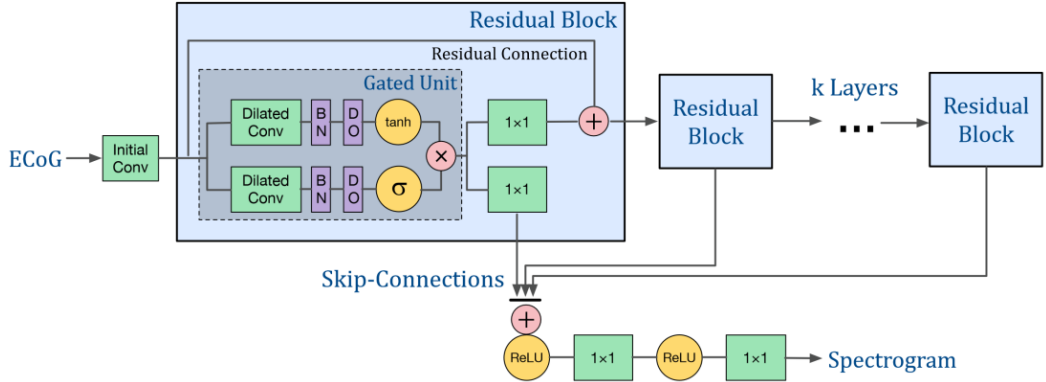

Figure 1. Network structure.

It has shown success in many audio generation applications [15], [17]–[22]. Here, we modified Wavenet to a regression model that takes time series ECoG signal input and outputs the spectrogram of speech. The structure of WaveNet is illustrated in Figure1. The model is based on a 1d convolution that filters along the time axis. The multiple electrodes of ECoG and frequency bands of spectrogram are treated as channels. After the initial convolutional layer, the model is constructed with several residual blocks. In each block, a gated unit is used as nonlinear activation to model the possible modulation effect of certain input temporal pattern on the output of the speech spectrogram. The signal then passes two branches of 1×1 convolution. The first branch is added to the input of the block as the residual. The second branch of all blocks are skipped and summed at the final post-processing layers that convert the extracted ECoG representation to the final spectrogram output.

Dilated convolution in the gated units contributes the most to capturing spectro-temporal patterns. This convolution has a perspective field that is larger than its filter length by skipping input values with steps of a certain dilation rate. With the same filter length, convolution with a larger dilation rate has a larger perspective field. By stacking residual blocks with dilated convolution layers of exponentially increasing dilation rate, the WaveNet is able to cover a large range of temporal samples with the number of parameters approximately equal to the logarithm of the temporal range. This significantly improves the parameter efficiency in terms of temporal coverage. Meanwhile, exponentially increasing dilation rate extracts a multi-scale representation of the ECoG signal. The summation of skip-connections from the residual blocks allows each block to only process residual information of a certain scale. Skip connections, along with residual connections, further improves parameter efficiency by decomposing the model to process multi-scale information in an incremental manner. To further prevent the model from overfitting, batch normalization and drop out are included in between each convolutional layer.

## III. EXPERIMENTS

### A. Data Acquisition

The brain activities were obtained from two patients with epilepsy and undergoing neurosurgery with an ECoG recording device [23]. The ECoG arrays have 8 by 8 electrodes with inter-electrode spacing of 4mm and sampling rate of 3051 Hz. Electrode arrays were implanted to cover the posterior lateral surface of the superior temporal gyrus. ECoG signals were recorded simultaneously when subjects were participating in short tasks within five minutes. During the task, both subjects were instructed to listen to speech audio (24 kHz sampling rate) of 50 different English words recorded by a native English female speaker. The first subject (S1) passively listened to each word. The same 50 words were repeated three times in different pseudo-random order. The second subject (S2) participated in two tasks with the same stimuli as S1. The subject was required to repeat the words they heard during the first task and translate the perceived words to Spanish during the second task. Each word was played twice for both tasks thus four ECoG signal examples of hearing the same word were recorded for S2. For the current work, only the response during the "listening" period is used to reconstruct the stimuli speech.

After synchronizing the speech waveform with ECoG signal by lagging speech with 168 ms behind, the speech spectrogram was generated by applying a 128 band-pass filter bank on the waveform. Center frequencies of filters are logarithmically spaced from 180-7000 Hz and have a bandwidth of 1/12 octave. The spectrogram is then subsampled to 32 bands in frequency and 100 Hz in time. ECoG signals were preprocessed with high gamma band-pass filter (70-150 Hz). The envelope of the filtered signal was then extracted by a Hilbert Huang transform and downsampled to 100 Hz to match the sampling rate of the

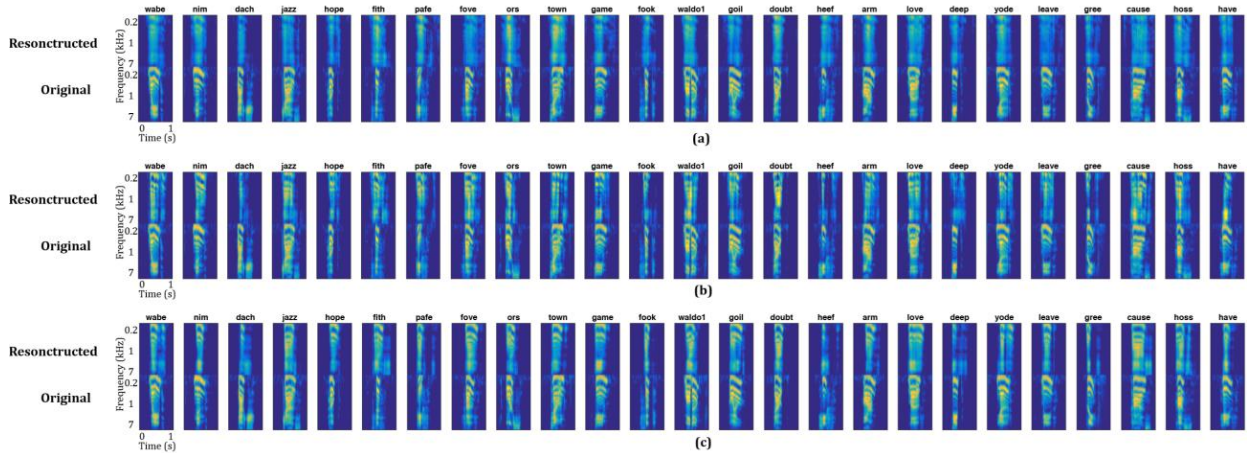

Figure 2. Reconstruction samples of (a) linear convolution, (b) Resnet, and (c) WaveNet

## B. Spectrogram reconstruction

For spectrogram reconstruction, we used a WaveNet with 10 residual blocks. Convolution with a filter width of 2 is used for dilated convolutions of each residual block. With the exponentially increasing dilation rate, the network covers 1240 ms temporal perspective field. As a result of grid search, Table. I and II show the optimized parameters for the network structure.

We also implemented linear convolution (equivalent to linear regression) and ResNet [24] as benchmark models. One-layer linear convolution has previously been used for correlating pSTG cortex responses and speech stimuli and for decoding speech from ECoG signals [11]. ResNet is widely used in various machine learning based audio signal processing tasks [25]–[31]. Linear convolution and ResNet are designed to have the same temporal perspective field width (1240ms) as the WaveNet. The linear convolution model uses one-layer of 1d convolution without activation function to transform ECoG signal with 64 channels to speech spectrogram with 32 channels. The ResNet model contains 8 residual blocks with filter length 4 and feature number 32. These settings are also optimized by hyper-parameter search.

We separate the datasets into training and testing set for k-fold (k=3 for S1 and k=4 for S2) cross-validation. Each partition contains 50 individual words in the testing set. The quantitative evaluation is based on the averaged testing result for each cross-validation. The datasets were segmented into short sequences of 1000 ms with overlapping 10 ms. Note that each word lasts around 400 ms and the spacing between words is about 1000 ms. Some of the short sequences contain only a single word, others have a partial word, and some contain only the silent period. 17k training sequences for S1 dataset and 26k for S2 dataset are included in each cross-validation partition. Despite a large number of segments, the actual words contained in the training set are limited (50 words, each repeated twice for S1 and 50 words, each repeated 3 times for S2)

Both mean squared error (MSE) and correlation coefficient (CC) are used as metrics to evaluate each method. CC is calculated based on the averaged correlation coefficient of all 32 spectrogram frequency channels. Table. III and IV report the testing result for subjects S1 and S2. Overall, WaveNet achieved the highest CC and lowest MSE for both datasets and deep models (WaveNet and ResNet) performed better than the linear convolution network. The number of trainable parameters for WaveNet, ResNet, and linear convolution was 509K, 132K, and 204K respectively. The fact that WaveNet obtained better results with more parameters suggests that the proposed WaveNet structure has a richer expressive power while having a better generalization capability with limited training data. Figure 2 illustrates reconstructed samples of each model. Linear convolution leads to over-smoothed spectrograms and fails to recover detailed spectral features. ResNet, on the contrary, has over-sharpened results and suffers from artificial spectro-temporal aliasing that causes discontinuity across time.

Table I. Hyperparameters for Wavenet structure

| conv layer | filter length | number of features |
| --- | --- | --- |
| initial conv | 32 | 16 |
| dilated conv | 2 | 32 |
| residual conv | 1 | 16 |
| skip conv | 1 | 32 |

Table II. Dilation rates for each residual block in the Wavenet

| residual block No. | 1 | 2 | 3 | 4 | 5 | 6 | 7 | 8 | 9 | 10 |
| --- | --- | --- | --- | --- | --- | --- | --- | --- | --- | --- |
| dilation rate | 1 | 2 | 4 | 8 | 16 | 1 | 2 | 4 | 8 | 16 |

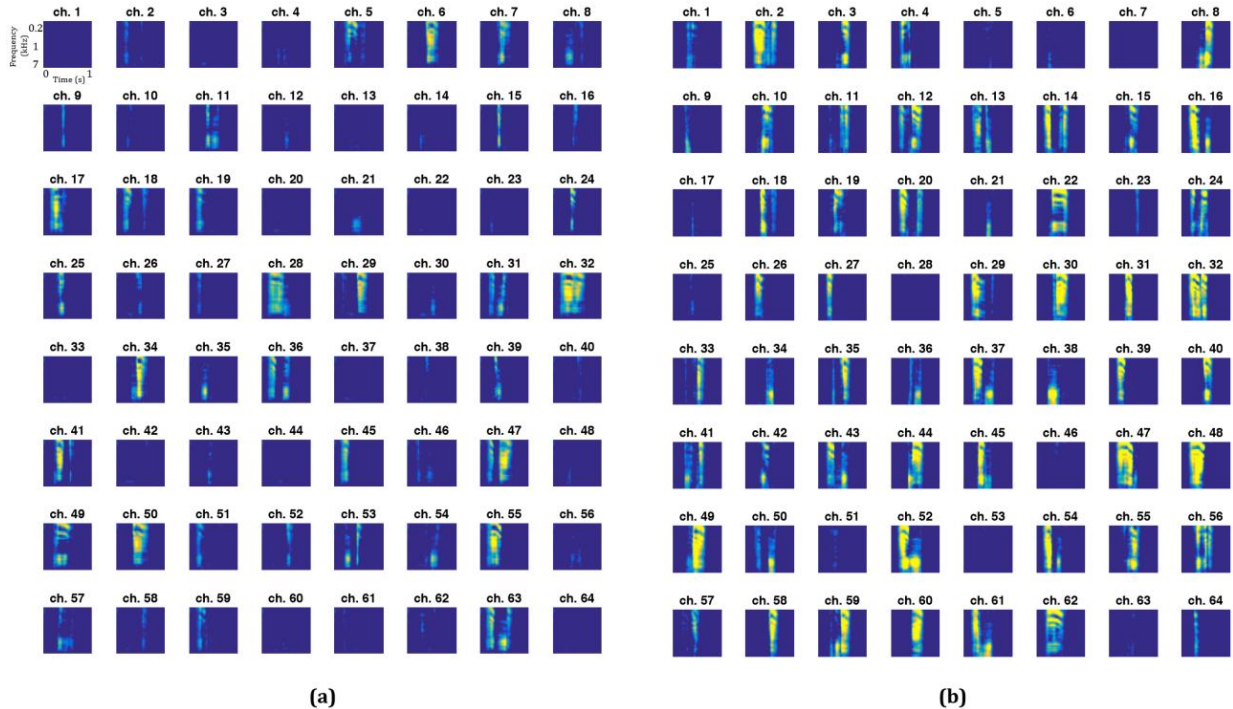

Figure 3. Impulse response for subject S1 (a) and S2 (b). Each subfigure corresponds to one ECoG electrode site.

Table III. Quantitive evaluation for S1 dataset

|     | linear conv | ResNet | WaveNet |
|-----|-------------|--------|---------|
| MSE | 0.86        | 0.79   | **0.68** |
| CC  | 0.57        | 0.61   | **0.65** |

Table IV. Quantitive evaluation for S2 dataset

|     | linear conv | ResNet | WaveNet |
|-----|-------------|--------|---------|
| MSE | 0.71        | 0.70   | **0.69** |
| CC  | 0.69        | 0.70   | **0.71** |

Even though the spectrograms of recovered and original words can be visually similar, to decode the word waveform intelligibly and correctly requires more precise reconstruction of the complex acoustic features, such as Formant frequencies and fast spectro-temporal fluctuation. We inverted the decoded spectrogram to a waveform using an iterative convex projection approach [32] for qualitative evaluation. We have found that, although the words in the testing set have already appeared in the training set, the ECoG recordings for the repeated words are quite different. There is no significant higher correlation between ECoG recordings of the same repeated words than of different words. Without sufficient data to uncover the complex non-linear relationship between ECoG signal and speech spectrogram, intelligible reconstruction of all speech stimuli is not expected. Nevertheless, several decoded words from WaveNet ("waldo", "yich" and "pave") are definitely intelligible. Even though some of the reconstructed spectrograms by ResNet looked quite similar to those by WaveNet, the reconstructed speech by ResNet is much less intelligible. The intelligible reconstruction further validates better expressive power and generalization ability of WaveNet than benchmark models.

Table V. Discovered phonetic features for ECoG electrodes on subject S1

| Electrode No.     | 7    | 8    | 50   | 54   | 55  |
|-------------------|------|------|------|------|-----|
| Perceived phoneme | /u:  | /ss  | /əu  | /sh  | /i  |

Table VI. Discovered phonetic features for ECoG electrodes on subject S2

| Electrode No.     | 38   | 39   | 40   | 47   | 48  |
|-------------------|------|------|------|------|-----|
| Perceived phoneme | /sh  | /i   | /i:  | /əu  | /u: |

*C. Phonetic Feature of Model Impulse Response*

Previous studies on pSTG cortex have shown selectivity to phonetic features for vowels and consonants [6], [11]. We were interested in evaluating whether our decoding approach also revealed similar phoneme selectivity for certain electrodes. We achieved this by investigating the "impulse response" for each electrode input. The impulse response for a given electrode can be considered as the stimuli speech (with its corresponding spectrogram with a certain temporal and spectral span) that causes a short-term response in the cortex area observed by this electrode. We generated an impulse signal for each of the 64 electrode channels. The impulse is generated between 500ms-600ms among the overall 1000ms duration. The impulse value is set to be the highest signal value of ECoG in our dataset. The rest of the temporal samples and electrodes are set to be zero. To generate the impulse response for a fitted model for each electrode, we drive the fitted model with the impulse input and obtain the decoded spectrogram for each electrode. Figure3 shows the impulse response for all 64 channels by training the model with all samples in the dataset. Since there exists an offset and orientation mismatch between ECoG arrays for different subjects, spatial distribution difference of impulse response for two models is expected. But the discovered phonetic features for both patients are robust and mostly consistent. We generated the "speach" signal from these impulse responses and were able to identify some of the phonemes, which are listed in Tables V and VI. We have found that the impulse responses for both subjects include the common set of /sh/, /i , /u: and /əu.

Previous studies investigating the pSTG area have revealed that certain electrode responses are sensitive to clusters of phonemes [6]. Here we further specified which phonemes are selected by certain electrodes. It is possible to uncover more electrodes with phonemes sensitivity given a bigger dataset of richer speech stimuli.

The consistency between WaveNet phonetic impulse response provides evidence for the phonetic selectivity of STG and suggests a promising direction for further study of the STG area with the discovered phonetic impulse responses.

## IV. CONCLUSION

In this work, we adapt WaveNet to decode speech from ECoG signals while listening to word stimuli. Despite a relatively small dataset, the high parameter efficiency of the adapted WaveNet is able to overcome the overfitting problem and generate reconstructed spectrograms with intelligible quality. The impulse response analysis of the fitted WaveNet model further reveals distinct phonetic encoding by some cortex areas. The phonetic features discovered by these impulse responses are consistent with previous discovers of STG area and suggest a promising direction for further analysis of the area and other auditory cortices.

## REFERENCES

[1] G. Hickok and D. Poeppel, "The cortical organization of speech processing," Nature Reviews Neuroscience, vol. 8, no. 5, p. 393, 2007.

[2] J. P. Rauschecker and S. K. Scott, "Maps and streams in the auditory cortex: nonhuman primates illuminate human speech processing,"Nature neuroscience, vol. 12, no. 6, p. 718, 2009.

[3] G. H. Recanzone and Y. E. Cohen, "Serial and parallel processing in the primate auditory cortex revisited," Behavioural brain research, vol. 206, no. 1, pp. 1–7, 2010.

[4] M. Steinschneider, "Unlocking the role of the superior temporal gyrus for speech sound categorization," Journal of Neurophysiology, vol. 105, no. 6, pp. 2631–2633, 2011.

[5] L. M. Romanski and B. B. Averbeck, "The primate cortical auditory system and neural representation of conspecific vocalizations," Annual review of neuroscience, vol. 32, pp. 315–346, 2009.

[6] N. Mesgarani, C. Cheung, K. Johnson, and E. F. Chang, "Phonetic feature encoding in human superior temporal gyrus," Science, vol. 343, no. 6174, pp. 1006–1010, 2014.

[7] J. R. Binder,J. A. Frost, T. A. Hammeke, P. S. Bellgowan, J.A. Springer, J. N. Kaufman, and E. T. Possing, "Human temporal lobe activation by speech and nonspeech sounds," Cerebral cortex, vol. 10, no. 5, pp. 512– 528, 2000.

[8] M. A. Howard, I. Volkov, R. Mirsky, P. Garell, M. Noh, M. Granner, H. Damasio, M. Steinschneider, R. Reale, J. Hind et al., "Auditory cortex on the human posterior superior temporal gyrus," Journal of Comparative Neurology, vol. 416, no. 1, pp. 79–92, 2000.

[9] I. DeWitt and J. P. Rauschecker, "Phoneme and word recognition in the auditory ventral stream," Proceedings of the National Academy of Sciences, vol. 109, no. 8, pp. E505–E514, 2012.

[10] P. W. Hullett, L. S. Hamilton, N. Mesgarani, C. E. Schreiner, and E. F. Chang, "Human superior temporal gyrus organization of spectrotemporal modulation tuning derived from speech stimuli," Journal of Neuroscience, vol. 36, no. 6, pp. 2014–2026, 2016.

[11] B. N. Pasley, S. V. David, N. Mesgarani, A. Flinker, S. A. Shamma, N. E. Crone, R. T. Knight, and E. F. Chang, "Reconstructing speech from human auditory cortex," PLoS biology, vol. 10, no. 1, p. e1001251, 2012.

[12] R. Wang, Y. Song, Y. Wang, and J. Viventi, "Long-term prediction of μecog signals with a spatio-temporal pyramid of adversarial convolutional networks," in Biomedical Imaging (ISBI 2018), 2018 IEEE 15th International Symposium on. IEEE, 2018, pp. 1313–1317.

[13] X. R. N. Wang, A. Farhadi, R. Rao, and B. Brunton, "Ajile movement prediction: Multimodal deep learning for natural human neural recordings and video," arXiv preprint arXiv:1709.05939, 2017.

[14] Z. Xie, O. Schwartz, and A. Prasad, "Decoding of finger trajectory from ecog using deep learning," Journal of neural engineering, vol. 15, no. 3, p. 036009, 2018

[15] A. Van Den Oord, S. Dieleman, H. Zen, K. Simonyan, O. Vinyals, A. Graves, N. Kalchbrenner, A. W. Senior, and K. Kavukcuoglu, "Wavenet: A generative model for raw audio." in SSW, 2016, p. 125.

[16] C. R. Holdgraf, J. W. Rieger, C. Micheli, S. Martin, R. T. Knight, and F. E. Theunissen, "Encoding and decoding models in cognitive electrophysiology," Frontiers in systems neuroscience, vol. 11, p. 61, 2017.

[17] J. Engel, C. Resnick, A. Roberts, S. Dieleman, D. Eck, K. Simonyan, and M. Norouzi, "Neural audio synthesis of musical notes with wavenet autoencoders," arXiv preprint arXiv:1704.01279, 2017.

[18] J. Shen, R. Pang, R. J. Weiss, M. Schuster, N. Jaitly, Z. Yang, Z. Chen, Y. Zhang, Y. Wang, R. Skerry-Ryan et al., "Natural tts synthesis by conditioning wavenet on mel spectrogram predictions," arXiv preprint arXiv:1712.05884, 2017.

[19] A. Tamamori, T. Hayashi, K. Kobayashi, K. Takeda, and T. Toda, "Speaker-dependent wavenet vocoder," in Proc. Interspeech, vol. 2017, 2017, pp. 1118–1122.

[20] K. Qian, Y. Zhang, S. Chang, X. Yang, D. Florˆencio, and M. HasegawaJohnson, "Speech enhancement using bayesian wavenet," in Proc. Interspeech, 2017, pp. 2013–2017.

[21] D. Rethage, J. Pons, and X. Serra, "A wavenet for speech denoising," arXiv preprint arXiv:1706.07162, 2017.

[22] A. v. d. Oord, Y. Li, I. Babuschkin, K. Simonyan, O. Vinyals, K. Kavukcuoglu, G. v. d. Driessche, E. Lockhart, L. C. Cobo, F. Stimberg et al., "Parallel wavenet: Fast high-fidelity speech synthesis," arXiv preprint arXiv:1711.10433, 2017.

[23] A. Flinker, E. Chang, N. Barbaro, M. Berger, and R. Knight, "Subcentimeter language organization in the human temporal lobe," Brain and language, vol. 117, no. 3, pp. 103–109, 2011.

[24] K. He, X. Zhang, S. Ren, and J. Sun, "Deep residual learning for image recognition," in Proceedings of the IEEE conference on computer vision and pattern recognition, 2016, pp. 770–778.

[25] S. Hershey, S. Chaudhuri, D. P. Ellis, J. F. Gemmeke, A. Jansen, R. C. Moore, M. Plakal, D. Platt, R. A. Saurous, B. Seybold et al., "Cnn architectures for large-scale audio classification," in Acoustics, Speech and Signal Processing (ICASSP), 2017 IEEE International Conference on. IEEE, 2017, pp. 131–135.

[26] D. Pollack, "Musical genre classification of audio," Master's thesis, Humboldt-Universit¨at zu Berlin, 2018.

[27] A. Jansen, M. Plakal, R. Pandya, D. P. Ellis, S. Hershey, J. Liu, R. C. Moore, and R. A. Saurous, "Towards learning semantic audio representations from unlabeled data," signal, vol. 2, no. 3, pp. 7–11, 2017.

[28] V. Ramanishka, A. Das, D. H. Park, S. Venugopalan, L. A. Hendricks, M. Rohrbach, and K. Saenko, "Multimodal video description," in Proceedings of the 2016 ACM on Multimedia Conference. ACM, 2016, pp. 1092–1096.

[29] S. Petridis, T. Stafylakis, P. Ma, F. Cai, G. Tzimiropoulos, and M. Pantic, "End-to-end audiovisual speech recognition," arXiv preprint arXiv:1802.06424, 2018.

[30] S. Oramas, O. Nieto, F. Barbieri, and X. Serra, "Multi-label music genre classification from audio, text, and images using deep features," arXiv preprint arXiv:1707.04916, 2017.

[31] A. Jansen, M. Plakal, R. Pandya, D. P. Ellis, S. Hershey, J. Liu, R. C. Moore, and R. A. Saurous, "Unsupervised learning of semantic audio representations," arXiv preprint arXiv:1711.02209, 2017.

[32] L. H. Arnal, A. Flinker, A. Kleinschmidt, A.-L. Giraud, and D. Poeppel, "Human screams occupy a privileged niche in the communication soundscape," Current Biology, vol. 25, no. 15, pp. 2051–2056, 2015.